\documentclass{aa}  
\usepackage{graphicx}
\usepackage{txfonts}
\usepackage{float}
\usepackage[usenames,dvipsnames]{color}
\usepackage[hyperfootnotes=false]{hyperref}
\hypersetup{
  colorlinks,
  citecolor=Violet,
  linkcolor=Blue, 
  urlcolor=Blue}
\usepackage{float}

\begin{document}

   \title{Observability of the Vertical Shear Instability in\\ protoplanetary disk CO kinematics}

   \author{Marcelo Barraza-Alfaro\inst{1}\fnmsep\thanks{Member of the International Max-Planck Research School for Astronomy and Cosmic Physics at the University of Heidelberg (IMPRS-HD), Germany}
      \and Mario Flock\inst{1}
      \and Sebastian Marino\inst{2,3} 
      \and Sebasti\'an P\'erez\inst{4,5} 
          }

   \institute{\inst{1}Max Planck Institute for Astronomy, Königstuhl 17, D-69117 Heidelberg, Germany\\
              \email{barraza@mpia.de}\\
              \inst{2}Institute of Astronomy, University of Cambridge, Madingley Road, Cambridge CB3 0HA, UK\\
              \inst{3}Jesus College, University of Cambridge, Jesus Lane, Cambridge CB5 8BL, UK\\
              \inst{4}Departamento de Física, Universidad de Santiago de Chile, Av. Ecuador 3493, Estación Central, Santiago, Chile\\
              \inst{5}Center for Interdisciplinary Research in Astrophysics and Space Exploration (CIRAS), Universidad de Santiago de Chile\\
             }

   \date{Received X; accepted Y}

 
  \abstract
   {Dynamical and turbulent motions of gas in a protoplanetary disk are crucial for their evolution and are thought to affect planet formation. Recent (sub-)millimeter observations show evidence of weak turbulence in the disk's outer regions. However, the detailed physical mechanism of turbulence in these outer regions remains uncertain. The vertical shear instability (VSI) is a promising candidate mechanism to produce turbulence in the outer parts of the disk.}
   {Our objective is to study the observability of the gas velocity structure produced by the vertical shear instability via CO kinematics with ALMA.}
   {We perform global 3D hydrodynamical simulations of an inviscid and locally isothermal VSI-unstable disk. We post-process the simulation results with radiative transfer calculations, and produce synthetic predictions of CO rotational emission lines. Following, we compute the line of sight velocity map, and its deviations from a sub-Keplerian equilibrium solution.
   We explore the detectability of the VSI by identifying kinematic signatures using realistic simulated observations using the CASA package.}
   {Our 3D hydrodynamical simulations of the VSI show the steady state dynamics of the gas in great detail. From the velocity structure we infer a turbulent stress value of $\alpha_{r\phi}=1.4 \times 10^{-4}$. On large scales, we observe clear velocity deviations in the order of 50 m s$^{-1}$ as axisymmetric rings with radially interspersed signs. By comparing synthetic observations at different inclinations we find optimal conditions at $i \lesssim 20^{\circ}$ to trace for the kinematic structures of the VSI. We found that current diagnostics to constrain gas turbulence from non-thermal broadening of the molecular line emission are not applicable to anisotropic VSI turbulence.}
   {We conclude that the detection of kinematic signatures produced by the vertical shear instability is possible with ALMA's current capabilities. Observations including an extended antenna configuration are required to resolve the structure (beam sizes below $\sim 10$ au). The highest spectral resolution available is needed ($\sim 0.05$ km s$^{-1}$ with ALMA Band 6) for robust detection. The characterization of the large-scale velocity perturbations is required to constrain the turbulence level produced by the VSI from gas observations.}


   \maketitle
   
%

\section{Introduction}


Understanding the origin of turbulence in protoplanetary disks is a key step towards the comprehension of how circumstellar disks evolve and how planets form. 
Globally, it can regulate the radial transport of angular momentum outwards through the generation of an effective viscosity, allowing the gas to accrete onto the central star (\citealt{Shakura1973, Lynden1974}; see also PPVI review chapter in \citealt{Turner2014}). Locally, turbulence can be crucial for different physical processes relevant for planet formation, for example, gas accretion onto protoplanets \citep{Picogna2008}, the formation and lifetime of vortices \citep{Fu2014}, dust settling \citep{Fromang2006}, and dust growth and concentration in pressure bumps \citep{Ormel2007, Birnstiel2010, Pinilla2012a, Pinilla2012b}. Moreover, radial variations in the strength of the turbulence can trigger the formation of pressure bumps \citep{Lyra2009, Regaly2012, Flock2015}.

First observations of disk turbulence from spatially unresolved line broadening with the  Submillimeter Array (SMA) were able to derive turbulence upper limits for the disks TW Hya, HD 163296 and DM Tau of around 40, 300 and 130 m s$^{-1}$ \citep{hug11,gui12}. Recently, the Atacama Large Millimeter/submillimeter Array (ALMA) has allowed a direct study of the gas turbulence in the disk's outer regions ($\gtrsim 30$ au), for the first time via spatially resolved observations of molecular gas lines. From the search of non-thermal line broadening, weak levels of turbulence ($\lesssim 5-10\%$ the local sound speed) have been constrained in  HD 163296, TW Hya, MWC 480 and V4046 Sgr protoplanetary disks \citep{Flaherty2015, Teague2016, Flaherty2017, Flaherty2018, Teague2018b, Flaherty2020}.
Most of the protoplanetary disk regions are expected to be weakly ionized \citep{Dzyurkevich2013, Desch2015}, therefore, the Magneto-Rotational-Instability \citep[MRI,][]{Balbus1991,Hawley1991} is unlikely to be active in most disk regions. In this MRI "dead-zone" pure hydrodynamical instabilities are expected to dominate as the turbulence source, where the prevailing mechanisms strongly depend on the local cooling timescale \citep[see e.g.][and references therein]{Lyra2019}. Dependent on the magnetic field orientation and strength, the dominant transport of angular moment in the dead-zone might be produced by a magnetic driven wind \citep{Turner2014,Bai2017}. However, these winds are typically launched at heights of around 5 to 7 scale heights above the midplane \citep{Gressel2020} and so at higher altitudes than the molecular line emitting region. 

The Vertical-Shear-Instability \citep[henceforth VSI,][]{Nelson2013} is a hydro-instability that operates in the outer disk regions with fast cooling \citep{Pfeil2019}, that produces anisotropic turbulence \citep{Stoll2017}. It arises naturally from the vertical gradient of the disk's angular velocity, where the surface layers rotate more slowly than the midplane. 
The VSI works more efficiently in isothermal disks \citep{Nelson2013}, but can also be active in a radiative disk \citep{Stoll2014, Flock2017,Flock2020}. Moreover, it can also be effective in a magnetized disk where non-ideal MHD effects suppress the MRI \citep{Cui2020}.
The VSI produces large-scale gas motions in its saturated state, which correspond to the long-wavelength modes or "body modes". The large-scale gas motions have a characteristic meridional circulation pattern, in which the vertical motions can produce stresses significantly larger than in the radial motions \citep{Stoll2017}. A small-scale component can also be seen as turbulent eddies \citep{Flores2020}.\\ 
The effective Shakura \& Sunyaev $\alpha$ viscosity generated in the radial direction by the VSI has been found to be small, with values between $10^{-4}$ and $10^{-3}$ \citep{Nelson2013, Flock2017, Stoll2014, Manger2018}. This low level of turbulence is consistent with the upper limits constrained from ALMA molecular gas line observations \citep{Flaherty2015, Teague2016, Flaherty2017, Flaherty2018, Flaherty2020} and with the recent estimates from CO evolution models \citep{Trapman2020}. Despite these observational constraints, no direct comparison between synthetic predictions and observations has been made.\\
Determining the turbulence level alone by unresolved observations is not enough to confirm the responsible mechanism. In order to advance in determining the dominant physical mechanism of the disk dynamics, observations that can resolve the spatial structure of the gas kinematics are necessary. Luckily, ALMA observations have allowed to directly probe the kinematics in disks and significant deviations from the expected sub-Keplerian motion in equilibrium \citep{Perez2015,Perez2018, Teague2018, Pinte2018, Perez2020, Casassus2019, Pinte2019, Teague2019, Pinte2020,DiskDynamics2020}, although most of the times these are interpreted as the dynamical imprint of Jupiter mass planets due to the lack of quantitative predictions from alternative scenarios.

In this work we study the observability of the VSI-triggered gas motions by producing synthetic ALMA observations of the CO gas line emission. We address which spatial and spectral resolutions are needed, and discuss how the identification of signatures of the VSI would add constraints to the disk properties. Especially the detection of large scale motions in gas kinematics may support or discard the VSI-origin of the gas turbulence.

The disk physical model, and description of the hydrodynamical simulations are presented in Section \ref{hydrosimulations}. The radiative transfer calculations and synthetic imaging are detailed in Section \ref{radiativetransfer}. In Section \ref{discussion} we discuss our main results. Finally, we summarize our work and draw our conclusions in Section \ref{conclusions}.

\section{Hydrodynamical Simulations}\label{hydrosimulations}

\subsection{Physical Model}\label{physicalmodel}

We adopted the physical model of the disk from \cite{Nelson2013}.
The physical model is described using cylindrical coordinates ($R$,$Z$,$\phi$). The hydrodynamical simulations described in Section \ref{hydrosimulations} are performed in spherical coordinates ($r$,$\theta$,$\phi$).

As initial conditions, we construct a set of 2D axisymmetric profiles for the density and rotation velocity that fulfill hydrostatic equilibrium.

The initial conditions for density and angular velocity are given by the equilibrium solutions:

\begin{equation}\label{eq:rho}
   \rho(R,Z) = \rho_0\,\left( \frac{R}{R_0}\right)^p\,\textrm{exp} \left( \frac{G\,M}{c_s^2} \left[\frac{1}{\sqrt{R^2+Z^2}}-\frac{1}{R}\right]\right),
\end{equation}

\begin{equation}\label{eq:omega}
   \Omega(R,Z) = \Omega_K\,\left[(p+q)\left(\frac{H}{R}\right)^2 + (1+q)-\frac{qR}{\sqrt{R^2+Z^2}} \right]^{1/2},
\end{equation}

\noindent where $\Omega_K$ is the Keplerian angular velocity $\Omega_K=\sqrt{GM/R^3}$, $H=c_s/\Omega_K$ is the local disk scale height, $c_s$ the local sound speed, $R_0$ is the reference radius, set to be equal to the code unit of length, and $p$ and $q$ are the power-law profiles for the density and temperature. $\rho_0$ sets the value of the gas midplane density at $r=R_0$.
The gas in our simulation is described by a locally isothermal equation of state, i.e. $P = \rho\, c_s^2$. The isothermal sound speed is proportional to the temperature, $c_s^2\propto T$. Therefore, $q$ also represents the radial power-law index of $c_s^2(R)$:
\begin{equation}
    c_s^2(R)=c_0^2\left(\frac{R}{R_0}\right)^q.
\end{equation}

\noindent The disk pressure scale height follows:
\begin{equation}
    H(R)=H_0\left(\frac{R}{R_0}\right)^{(q+3)/2},
\end{equation}
where $H_0$ is the disc scale height at the reference radius $R_0$, set such that $H_0/R_0=0.1$. 
In our simulations the midplane density slope is $p=-1.0$ , while the temperature slope is $q=-0.5$, resulting in a flared disk with a flaring index of $0.25$. 

Looking at the last term of the prescription given in equation \ref{eq:omega}, we observe that for the same cylindrical radius the upper layers of the disk have slower angular velocity than the midplane, i.e., the disk has a vertical shear, which sets the instability in our simulations.

\subsection{Numerical method}\label{methodhydrosimulations}

We run global hydrodynamical simulations of an inviscid disk unstable to the Vertical Shear Instability.
The simulations are performed with the public version of the Godunov-grid-based code  \textsc{PLUTO\footnote{\url{http://plutocode.ph.unito.it/}}} \citep[version 4.3][]{Mignone2007}. \\
We use the HD module of the PLUTO code to solve the Navier-Stokes equations of classical fluid dynamics:
\begin{equation}
    \frac{\partial \rho}{\partial t}+\vec{\nabla}\cdot (\rho\vec{v})=0
\end{equation}
\begin{equation}
    \frac{\partial (\rho \vec{v})}{\partial t}+\vec{\nabla}\cdot (\rho\, \vec{v}\, \vec{v}^\text{T})=-\vec{\nabla}P - \rho \vec{\nabla}\Phi, 
\end{equation}
were $\rho$ is the gas mass density, $\vec{v}$ is the gas velocity vector, $P$ is the pressure, and $\Phi$ is the gravitational potential. The fluid is affected by the gravitational potentials of a star only, $\Phi_{\star}= -G M_{\star}/r$. 
The hydrodynamical equations were solved using a second-order accurate scheme with linear spatial reconstruction (LINEAR), second-order Runge-Kutta time-stepping, and the Harten-Lax-van Leer-Contact (HLLC) Riemann solver. The Courant number is set to 0.25.\\

The computational domain extends from $0.4$ to $2.5$ code units of length in the radial direction. In the vertical direction the model covers $\sim 9$ disk scale heights in total, $\sim 4.5$ at each side of the equator. In the azimuthal direction the grid covers the full $2\pi$ rad. 
The spherical grid is logarithmically spaced in the radial direction $r$, while evenly-spaced in colatitude and azimuth. The resolution of the grid is $(r,\theta,\phi)=(512,128,1024)$, which gives a resolution of 14 cells per scale height and cells with an aspect ratio $\delta r{:}r\delta\theta{:}r\delta\phi$ of approximately $1{:}1.9{:}1.7$ .
Since our simulation follows a locally isothermal equation of state, the simulation's physical radius can be re-scaled. However, the reference aspect ratio of $H_0/R_0= 0.1$ set in our simulations fits best for a T Tauri disk model re-scaled to around 100 au. We display a summary of parameters used in the simulations in the upper section of Table \ref{tab:simparams}.

We adopted similar boundary conditions as \cite{Flock2017}.
In colatitude, it corresponds to a modified outflow boundary condition which enforces zero inflow of gas into the computational domain. In the density, it extrapolates the logarithmic field along the meridional direction into the ghost cells. Additionally, we include a softening of the azimuthal velocity radial gradient, only at the meridional interface between the ghost cells and the computational domain to reduce effects from the boundary corners. In the radial direction, enforced zero inflow of gas is imposed. A buffer zone is also applied close to the inner and outer radial boundaries, whose sizes are equal to $20\%$ of the inner and outer edge radius, respectively. In the buffer zones, the density and radial velocity are damped to the initial profiles.

\subsection{Simulation results}
\begin{figure}[H]
    \centering
    \includegraphics[angle=0,width=1.0\linewidth]{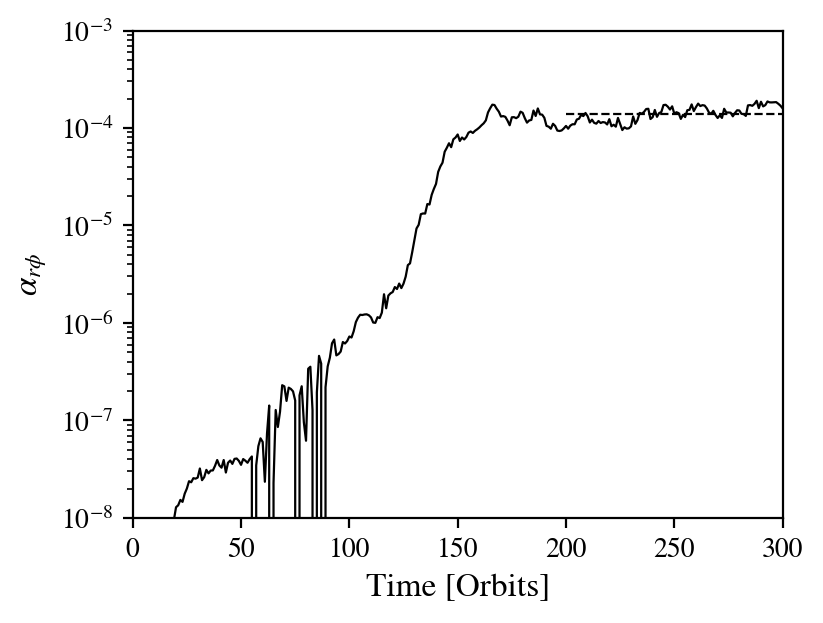}
    \caption{Time evolution of the stress-to-pressure ratio $\alpha_{r\phi}$ of the VSI-unstable 3D hydrodynamical simulation. We show the time evolution in units of the orbital timescale at $R=100 \rm au$. The dashed line indicates the time-averaged value of $\alpha_{r\phi}$ between $200$ and $300$ orbits.}
    \label{fighydro_alpha}
\end{figure}

\begin{figure}[H]
\centering
\includegraphics[angle=0,width=1.0\linewidth]{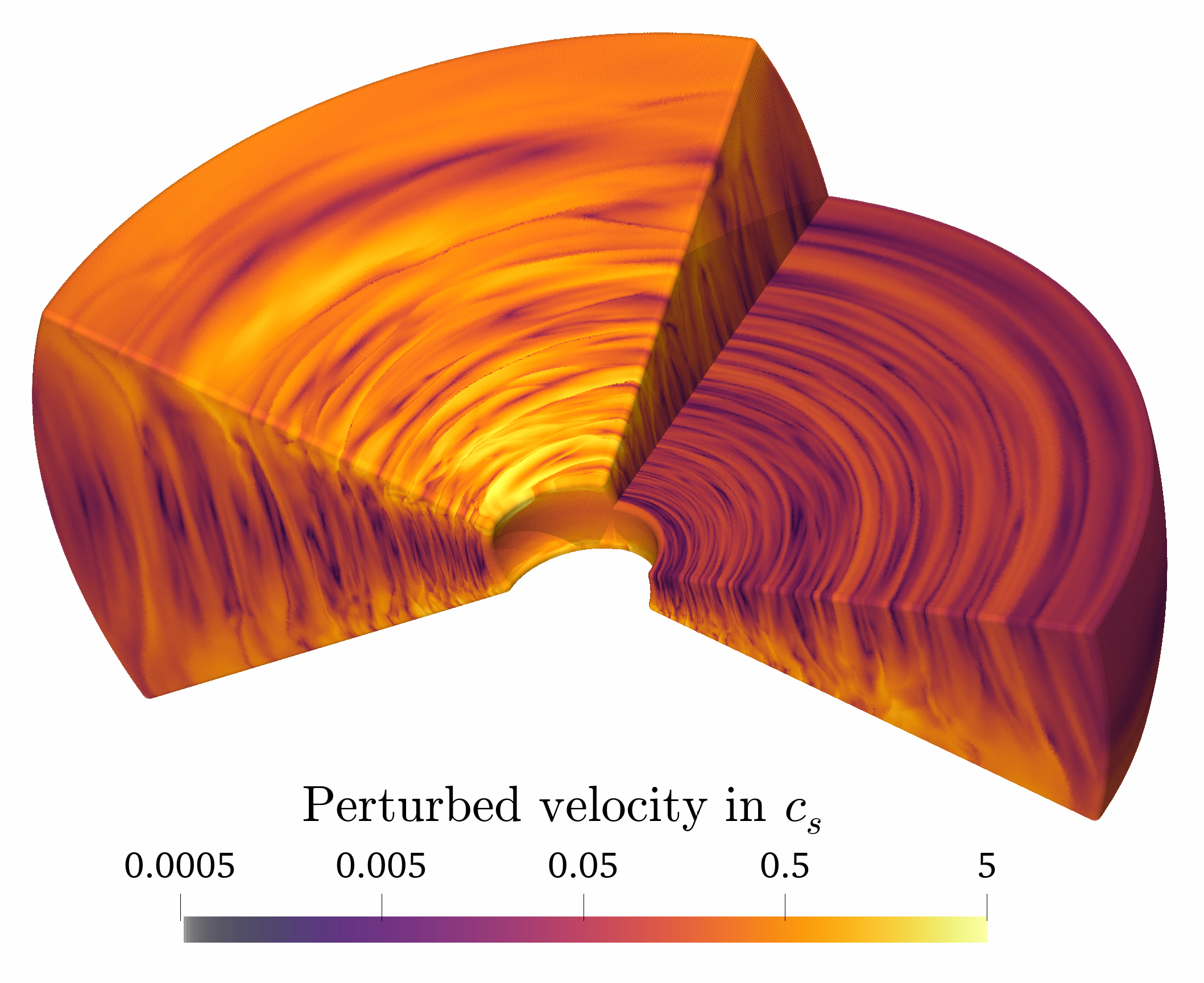}
\caption{3D rendering of a VSI-unstable protoplanetary disk. The color map shows the perturbed total velocity of the gas in units of the local sound speed. Only half of our simulation domain in azimuth is shown for visualization purposes.}
\label{fig3d}
\end{figure}

Fig.~\ref{fighydro_alpha} shows the temporal evolution of the $\alpha_{r\phi}$ value consisting of the radial Reynolds stress-to-pressure ratio. We determine $\alpha_{r\phi}$ using the pressure-weighted stress-to-pressure ratio
\begin{equation}
\rm \alpha_{r\phi} = \frac{ \int T_{r \phi} dV} {\int P dV} = \frac{ \int \rho
  v'_{\phi}v'_{r}dV} {\int \rho c_s^2 dV},
\label{eq:ALPHA}
\end{equation}
with the volume $dV$, and $v'_{\phi}$ and  $v'_{r}$ representing the turbulent components of the velocities are determined by subtracting the azimuthally averaged value at each r and $\theta$. Fig.~\ref{fighydro_alpha} shows the time evolution, with a growth from very small values (we set the initial radial and vertical velocities to zero) until reaching a saturated value after around 170 orbits of evolution with $\alpha_{r\phi}\sim 1.4 \times 10^{-4}$. For the further post-processing we use a snapshot after 300 orbits. A 3D plot showing the 3D gas dynamics of the disk is shown in Fig.~\ref{fig3d}.

\begin{figure*}[htp!]
\centering
\includegraphics[angle=0,width=1.0\linewidth]{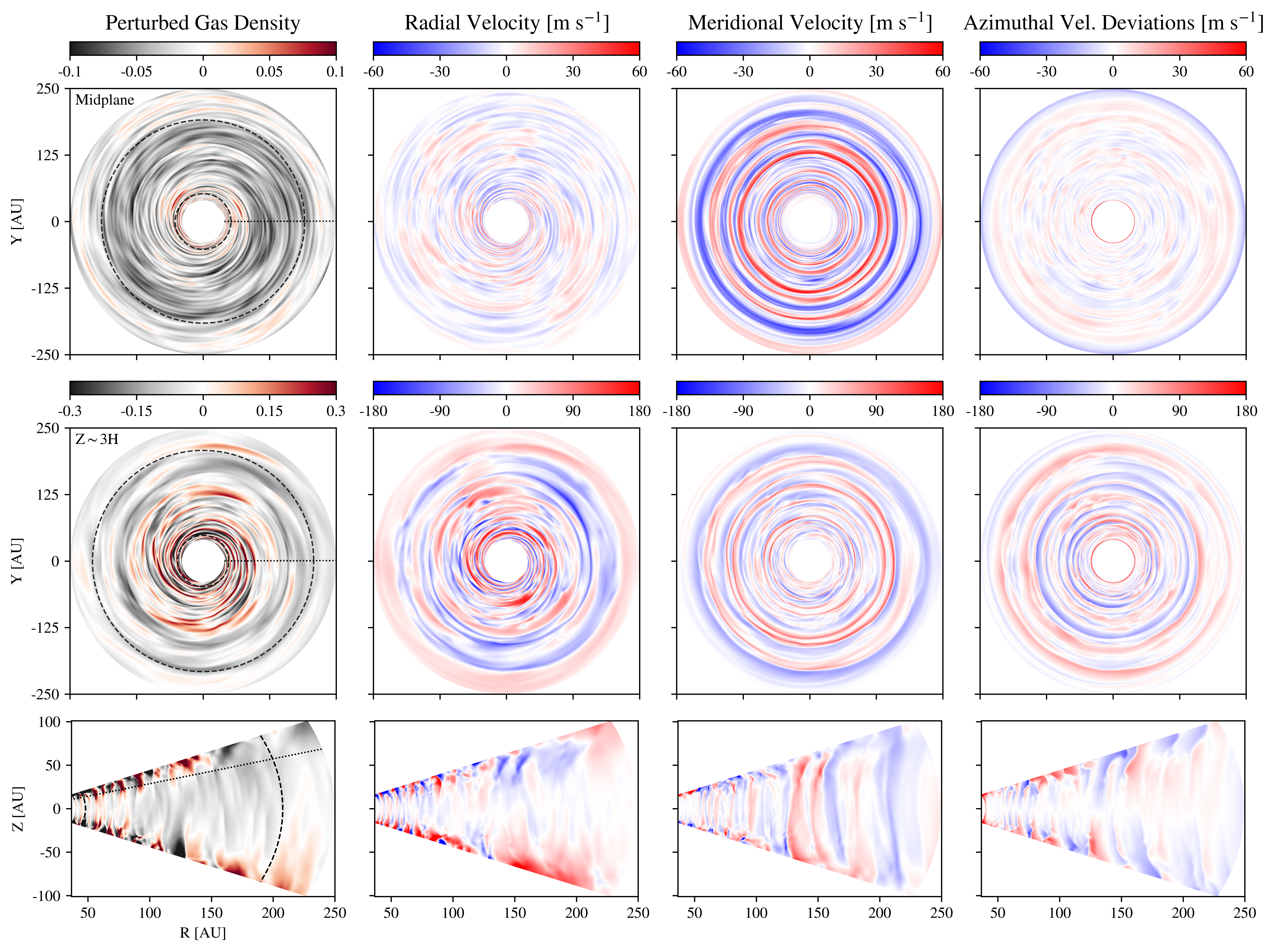}
\caption{Results of the 3D hydrodynamical simulation. The panels show snapshots of the density and the velocities, taken after $300$ orbital periods. \emph{From left to right}: Perturbed gas density, radial velocity, meridional velocity, and azimuthal velocity deviations from a sub-Keplerian velocity equilibrium solution. The velocity values are in physical units, where the simulation code unit of length is re-scaled to 100 au, and the central star is a Solar-mass star.
\emph{From top to bottom}: $r$-$\phi$ slices of the fields at the disk midplane, $r$-$\phi$ slices of the fields at approximately three pressure scale heights ($3H$) above the midplane, and $r$-$\theta$ slices of the fields.
In the left-most column plots, the circles in dashed lines indicate the location of the inner and outer buffer zones. The dotted lines in the $r$-$\phi$ slices indicate the azimuth where the $r$-$\theta$ slice is taken, while in the $r$-$\theta$ plots the dotted line indicates the location where $z \sim 3H$.}
\label{fighydrosimus_all}
\end{figure*}

As a next step, we present detailed horizontal and vertical 2D plots out of the 3D dataset using the same snapshots after $300$ orbits (measured at the code unit of length). These 2D plots are shown in Figure \ref{fighydrosimus_all}. The fields have been computed re-scaling the code unit of length to 100 au, and considering a central Solar-mass star, therefore, the snapshot corresponds to a disk evolution time of $\sim$0.3 Myr. Each row represents snapshots of the midplane, top row, at three scale heights middle row, and a vertical cut in the R-Z plane. The four columns present density and the three velocity components. In the left column of Fig.~\ref{fighydrosimus_all}, we plot the gas density perturbations, $(\rho-\rho_0)/\rho_0$. At the midplane and upper layers we see a turbulent structure with spirals and ring-like structures. At the midplane the density deviations are on a level of several percent of the background value. At the upper layers of the disk the density deviations reaches levels of up to 30\%. We note that even though the fluctuations are strong in the upper layers, there is no particular morphology that could be easily identified as VSI-generated.

Figure \ref{fighydrosimus_all}, right columns, show the 3D velocity structure typical of a disk in which the VSI is active in its saturated state. 
Strong perturbations are present in the three velocity components.
In the radial direction, the velocity shows turbulent behavior, thus, not well-defined structures are seen. At the midplane the radial velocities reach values of several 10's of meters per second, however, we note that in the upper layers these radial motions can reach over 100's of meters per second. Such large values become of interest especially when looking later on the line emission.\\
Similarly, very turbulent structures are present in the azimuthal velocity deviations from a disk in sub-Keplerian rotation (expected from an unperturbed disk) at the midplane. Close to the surface, however, axisymmetric rings are visible. Interestingly, in some of these rings the rotation is sub-Keplerian in one hemisphere while super-Keplerian in the other. We also note that similar to the radial perturbations, the azimuthal ones are also weakest in the midplane with values of around 10's meters per second while they become larger at the upper layers.\\
We summarize that the density perturbations, radial velocity perturbations and the azimuthal velocity deviation from a disk in Keplerian rotation are weak close to the disk midplane, however, above three pressure scale heights they can reach up to 100's of meters per second.\\ 
Even more interesting are the meridional velocity motions, seen in Fig.~\ref{fighydrosimus_all} third column. There are prominent large scale motions through the whole vertical extent of the disk reaching velocities of around 50 meters per second. These motions are mostly axisymmetric, and therefore they show a clearer signature compared to the radial and azimuthal components.

\section{Synthetic observations}\label{radiativetransfer}

To produce synthetic images of molecular line emission of a VSI-active disk, we post-process the outputs of our simulations with the Monte-Carlo radiative transfer code \textsc{radmc-3d\footnote{\url{http://www.ita.uni-heidelberg.de/~dullemond/software/radmc-3d}}} \citep[version 0.41,][]{Dullemond2012}. We build our \textsc{radmc-3d} input files partly based on the \textsc{fargo2radmc3d\footnote{\url{https://github.com/charango/fargo2radmc3d}}} script \citep{Baruteau2019}. 

We explore as observable the spatially resolved velocity centroid map (also labeled line of sight velocity map or Moment one map) computed from synthetic CO channel maps. To study the optically thin and optically thick cases, we compute the synthetic maps for three different CO isotopologues, $^{12}$CO, $^{13}$CO and C$^{18}$O.
We chose the J=2-1 transition for the three isotopologues, centered at $\sim 230.538$ GHz for $^{12}$CO,  $\sim 220.399$ GHz for $^{13}$CO, and $\sim 219.560$ GHz for C$^{18}$O. All three transitions are observable with ALMA Band 6.

\subsection{Setup}\label{rtsetup}

As mentioned before, we can re-scale the simulation since it follows an isothermal equation of state. We assume a central star of $1 M_{\odot}$, and re-scale our density and velocity fields of the hydrodynamical simulation to $R_0=100$ au, and use it as input into \textsc{radmc-3d}. To improve the performance, we volume-average the simulation data onto a coarser grid, halving the grid resolution for the radiative transfer post-processing. We also extended our disk, including an inner disk that goes from 10 au to the simulation grid's inner edge of 40 au, that follows the equilibrium solution used as the initial condition in the simulation (see equations \ref{eq:rho} and \ref{eq:omega}). Therefore, our disk model extends from 10 to 250 au. Additionally, we set the gas density of the model so the total gas mass of the disk in molecular hydrogen is $0.05\,M_{\odot}$, which result in a gas surface density of $\sim 3$ g cm$^{-2}$ at 100 au.

We compute the dust temperature via thermal Monte Carlo and assume that gas and dust have the same temperature. We adopt a constant gas-to-dust mass ratio of $100$. The dust is composed of $60\,\%$ astrosilicates \citep{Draine1984} and $40\,\%$ amporphous carbon \citep{Li1997}. The optical constant of the mixture was calculated using the Bruggeman mixing formula, with a resulting intrinsic density of $2.7\,\rm g\,cm^{-3}$. Mie theory was used to compute the dust opacities \citep{Bohren1983}. In our model the dust size distribution is defined adopting 10 logarithmically spaced dust size bins, with sizes ranging between a minimum of $0.01\,\mu\textrm{m}$, and a maximum of $1\,\textrm{cm}$. The dust size distribution follows a power-law profile $n(s)\propto s^{-p}$ with $p=3.5$, in which the mass in the i-th size bin is $M_{d,i}=\chi M_{gas}\times s_i^{4-p}/\sum_i s_i^{4-p}$, with $s_i$ the size of the i-th dust bin, $\chi=0.01$ the dust-to-gas mass ratio and $M_{gas} = 0.05\,M_{\odot}$ the gas mass in the disk model \citep[see also][]{Baruteau2019}. We adopted a simplified dust vertical settling prescription in which the dust smaller than $10\,\mu\textrm{m}$ has the same scale height as the gas, whereas the dust with sizes larger than $10\,\mu\textrm{m}$ is settled towards the midplane, with the dust scale height of the i-th dust size bin following: $H_{d,i}=\Lambda H_{g}(s_i/s_{min})^{-0.1}$, where $H_{g}$ is the scale height of the gas and $s_{min}$ is the minimum dust size in the model. The settling parameter $\Lambda$ is set to $0.2$, selected to roughly match our mm-sized dust scale height with constraints from observations \citep[e.g.][]{Pinte2016, Villenave2020}. The adopted dust scale heights are comparable to the obtained using standard diffusion models \citep[e.g.][]{Dubrulle1995} for the disk-averaged vertical alpha in our simulation $\alpha_{z\phi}\sim 5.1\times 10^{-4}$, computed for one hemisphere of the disk following Eq. 4 in \citealt{Stoll2017}. Global simulations of the VSI that including dust and gas dynamics constrain larger dust scale heights due to the vertical mixing produced by the VSI vertical motions \citep{Stoll2016, Flock2017, Flock2020}. Nevertheless, in our models a different vertical distribution of dust would only change the vertical temperature gradient of the disk in our radiative transfer predictions, which would have a small effect on the observable kinematic signatures.\\
The central object surface temperature is $T_*=7000\,\textrm{K}$ with a radius of $1\,R_{\odot}$, and we assume that it emits as a perfect Black-body. The obtained midplane temperature approximately matches the used in the hydrodynamical simulations, which follows a power law $\propto R^{-0.5}$, with R the cylindrical radius. Given the assumed central star, the midplane temperature goes from $\sim 40$ K at 40 au to $\sim 16$ K at 250 au. 
The dust is not included in the image ray-tracing to avoid optical depth effects produced by the dust continuum in the line emission. We use $10^9$ photon packages to compute the dust temperature via thermal Monte Carlo including absorption only, while $10^8$ photon packages for the image ray-tracing.
A summary of relevant parameters used in the radiative transfer calculations are shown in Table \ref{tab:simparams}.

For the molecular abundances, we assume a constant fraction of $^{12}$CO relative to H$_2$ of $1\times10^{-4}$. To obtain the abundance of $^{13}$CO and C$^{18}$O isotopologues, we scale the $^{12}$CO abundance with the ISM isotope ratios   [$^{12}$C]/[$^{13}$C$]\sim77$ and
[$^{16}$O]/[$^{18}$O]$\sim560$, respectively \citep{Wilson1994}.
The line emission is computed assuming local-thermodynamic-equilibrium (LTE). The molecular data is from the \textsc{LAMDA\footnote{\url{https://home.strw.leidenuniv.nl/~moldata/}}} database \citep{Schoier2005}. Variations of CO abundance from freeze-out or photo-dissociation are not included in our model. Although we omit the effect of CO freeze-out, the dust temperature obtained via thermal Monte Carlo shows a vertical gradient, leading to stronger emission in the upper layers. 
Using the molecular abundance fraction above, we generate sets of 3D grids of CO number density, and the three components of the velocity. Finally, the gas temperature is derived using the dust temperature from RADMC3D (see above).\\

As the gas motions produced by the VSI are roughly axisymmetric, we only study the disk emission by varying the inclination of the disk (5$^{\circ}$,45$^{\circ}$ and 85$^{\circ}$), while keeping the position angle (PA) fixed to 90$^{\circ}$ (East of North). We assume a distance to the source of 100 pc. The disk near side is the South, and its rotation is counter-clockwise with respect to the observer. The total integrated flux of the $^{12}$CO$(2-1)$, $^{13}$CO$(2-1)$ and C$^{18}$O$(2-1)$ synthetic datacubes are approximately 42, 23 and 13 Jy km s$^{-1}$, respectively.

We compute the synthetic cubes to have a total bandwidth of 6 km s$^{-1}$, with $600$ channels of $0.01$ km s$^{-1}$. Then, we average these to obtain data cubes with a resolution of $0.05$ km s$^{-1}$, i.e., the current maximum resolution of ALMA, that we use in the analysis. For a first approximation of a synthetic observation, we convolved the raw resulting synthetic channel maps by a circular-Gaussian beam, and study the addition of noise following the white noise model implemented in \textsc{fargo2radmc3d} (see also \citealt{Baruteau2019}).
A mask of $0.42$ arcseconds ($42$ au) was applied to all raw images to remove the emission from the disk regions close to the simulation's inner edge. Additional data cubes varying the velocity resolution and spatial resolution are computed to explore the dependency of our results with resolution which we discuss in Section \ref{Deviations}.
The emission of four different velocity channels convolved by a circular Gaussian beam of 50 mas (5 au) is shown in Figure \ref{channelmaps}, for each disk inclination. It is readily seen that the VSI produces velocity perturbations in individual channels, seen as a corrugated pattern at the edges of the emission, more prominent with decreasing disk inclination.

\subsection{ALMA synthetic observations}\label{simulatedobs}

For a more accurate prediction of how our model would look in an interferometric ALMA observation, we simulate observations with CASA \citep{McMullin2007} version 5.6 using the task \texttt{simobserve}.
We use our $^{12}$CO$(2-1)$ synthetic channel maps as input to \texttt{simobserve}. We simulate a 20 h integration in configuration C43-8 (8.5 km longest baseline) combined with a 4.4 h integration in C43-5 (1.4 km longest baseline). As described in the ALMA proposers guide, this configuration is ideal for a good uv-coverage in both short and long baselines.

We compute model visibilities using \texttt{simobserve} for each of the disk inclinations, with channel velocity resolution of 0.05 km s$^{-1}$.  Then, we corrupt the visibilities with the task \textsc{ms.corrupt()} to obtain a RMS of $\sim 1.5$ mJy beam$^{-1}$ in each channel. This is roughly the expected RMS for a precipitable water vapour (PWV) of 0.9 mm and 20 h of integration time according to the ALMA sensitivity calculator. Finally, we compute CLEANed spectral cubes with CASA \texttt{tclean}. The restored beam has a FWHM of $\sim 80\times 60$ mas (Briggs weighting 1.0). The average signal-to-noise per beam and channel is $\sim10$ at $\sim100$ au from the central star, and larger than 5 in all the disk.

\subsection{Velocity centroid maps and best fit sub-Keplerian disk model}\label{observables}

We compute velocity centroid maps (v$_0$) from the ALMA simulated data cubes using a Gaussian function to fit the line emission in each pixel. For this we use the python package \textsc{bettermoments\footnote{\url{https://github.com/richteague/bettermoments}}} \citep{Teague2018c, Teague2019b}. Alternatively, fitting a quadratic function to the line centroid could produce more precise results, however, a very high average signal-to-noise observation is needed for the method to work properly. 

Following, we extract the velocity perturbations of the projected velocity image. For this purpose we use the Extracting Disk DYnamics python suite \textsc{eddy\footnote{\url{https://github.com/richteague/eddy}}} \citep{eddy} to obtain the sub-Keplerian disk model that best fit our $^{12}$CO$(2-1)$ synthetic velocity centroid maps. 

The model assumes that the disk emitting surface is parametrized by:
\begin{equation}
   z(r) = z_0 \times \left(\frac{R}{1^{\prime\prime}}\right)^{\psi} \times \exp\left(-\left[\frac{R}{R_{\rm taper}}\right]^{q_{\rm taper}}\right),
\end{equation}

where $z_0$ and $R_{\rm taper}$ are in arcseconds. The disk rotation curve and radial velocity follow a power-law profile:
\begin{equation}
    v_{\phi} = v_{\phi,100}\times\left(\frac{R}{100\,\textrm{au}}\right)^{v_{\phi,q}}
\end{equation}
\begin{equation}
    v_{R} = v_{R,100}\times\left(\frac{R}{100 \,\textrm{au}}\right)^{v_{R,q}},
\end{equation}
with $R$ the cylindrical radius, and $v_{R,100}$ and $v_{\phi,100}$ the disk radial and azimuthal velocity at $100\,\textrm{au}$ from the star. 
The disk velocity model projected into the line of sight considers the contributions of both radial and azimuthal velocity components:
\begin{equation}
    \textrm{v$_{mod}$} = v_{\phi}\cdot \cos{\phi} \cdot \sin{i} +  v_{R}\cdot \sin{\phi} \cdot \sin{i} +v_{LSR},
\end{equation}
\noindent where $\phi$ is the polar angle of the pixel (measured east of north relative to the red-shifted major axis) and $v_{\rm LSR}$ is the systemic velocity. 

We set as free disk parameters the disk position angle, systemic velocity, disk center, the radial and azimuthal velocity at $100$ au from the star, the slope of the radial and azimuthal velocity power-law profiles, and emission surface parameters ($z_0$, $\psi$, $R_{\rm taper}$ and $q_{\rm taper}$). We fix the distance to the source to $100$ pc and the disk inclination equal to the value used to compute the input data cube.

\begin{figure*}[htp!]
\centering
\includegraphics[angle=0,width=1.0\linewidth]{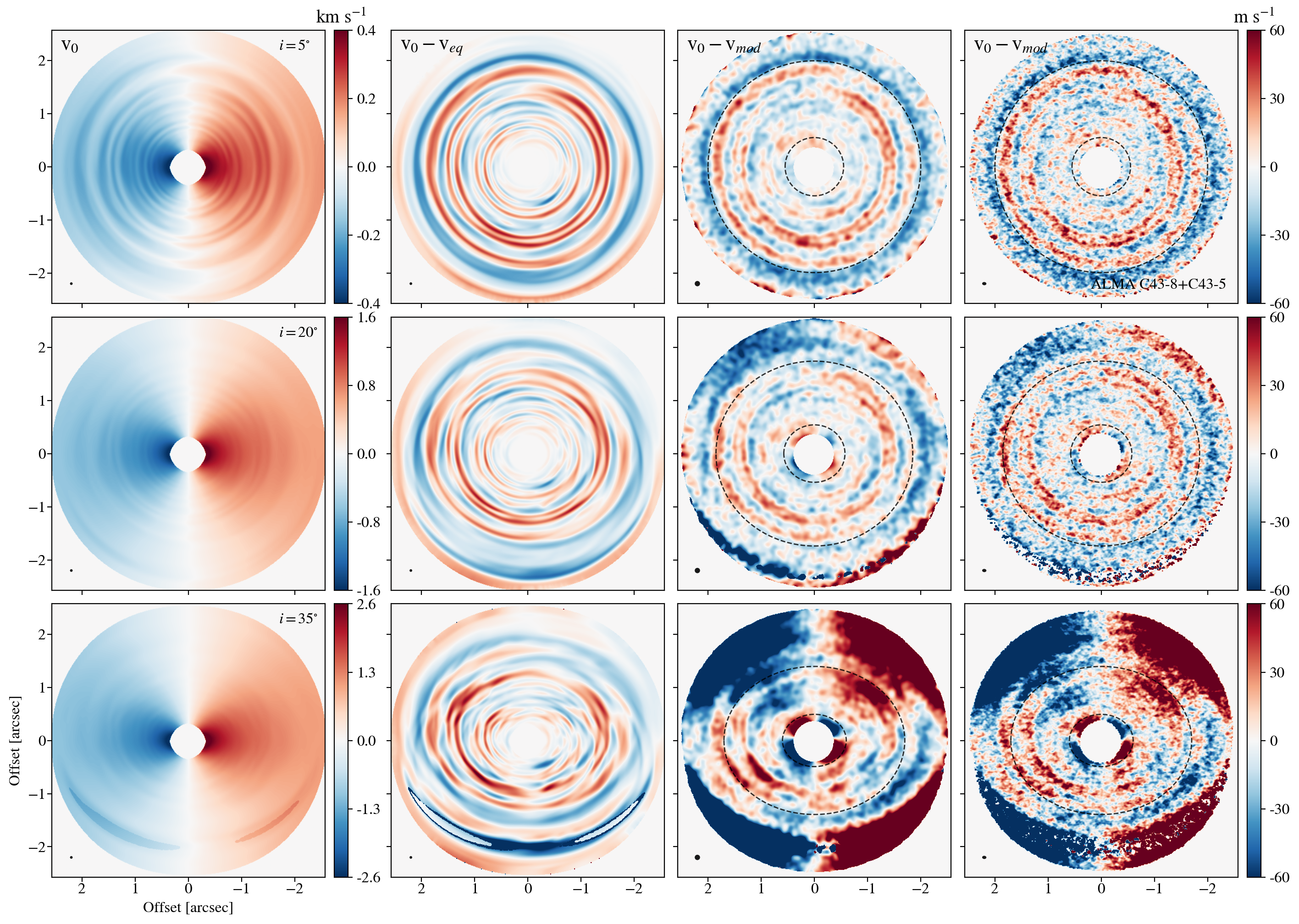}

\caption{Results of the line of sight velocity map and extracted velocity perturbations from a VSI unstable disk $^{12}$CO(2-1) synthetic lines observations. The velocity centroid of the line was computed at each pixel from mock data cubes with a velocity resolution of 0.05 km s$^{-1}$.  
The input fields are shown in Figure \ref{fighydrosimus_all}, which corresponds to a disk after $\sim 0.3$ Myr of evolution. From top to bottom, are shown the results for disk inclinations of $5^{\circ}$, $20^{\circ}$ and $35^{\circ}$.
\emph{First column:} Velocity centroid maps ($v_0$). The images are convolved by a circular Gaussian beam of 50 mas and have no noise. 
\emph{Second column:} Residual map of subtracting to $v_0$ the velocity centroid map obtained from a disk following an equilibrium solution ($v_{eq}$).
\emph{Third column:} Residual of subtracting to $v_0$ the best fit disk model obtained using \textsc{eddy} ($v_{mod})$. The model corresponds to a geometrically thick disk, that follows power-law profiles for the velocities in the radial and azimuthal directions. The images in this case are convolved by a 0.1 arcseconds circular Gaussian beam and have an RMS noise of $\sim 1.5$ mJy beam$^{-1}$.
\emph{Fourth column:} The same as the third column, but for a 20 h Cycle 7 ALMA simulated observation using configurations C43-8 and C43-5, with an RMS noise of $\sim 1.5$ mJy beam$^{-1}$. The resulting beam size of our synthetic observation is $80\times60$ arcseconds with a PA of -86.8$^{\circ}$. The beam size is shown with a black circle at the bottom left of each panel. The black-dotted ellipses in panels of columns 3 and 4 are the inner and outer edge of the region considered to obtain the best fit model.
The x- and y- axes indicate the R.A. and Dec. angular offset from the central star's position in arcseconds.}
\label{rtfigure}
\end{figure*}

A series of MCMC chains are run to find the best fit sub-Keplerian model for a geometrically thick disk. We use 128 walkers that take 2000 burn-in steps and additional 500 steps to sample the posterior distributions for the model parameters. A delimited radial region of the disk is considered in the model fitting, set to [0.55,2.0], [0.58,1.85] and [0.6,1.7] arcseconds for inclinations of 5, 25 and 35 degrees, respectively. Finally, we extract the velocity perturbations subtracting the velocity centroid map of the best fit disk model (v$_{mod}$) to the original (v$_0$). The velocity deviations results are presented in Figure \ref{rtfigure}, and discussed in the following section.

\subsection{Deviations from a sub-Keplerian disk model}\label{Deviations}

The line of sight velocity maps, and extracted velocity deviations from a sub-Keplerian disk for three different disk inclinations ($5^{\circ}$, $20^{\circ}$ and $35^{\circ}$) are shown in Figure \ref{rtfigure}.
The maps are computed for line emission data cubes with a velocity resolution of 0.05 km s$^{-1}$.
In the first column, the line centroid at each pixel is presented as the disk velocity in the line of sight (v$_0$), for images convolved by a circular Gaussian beam of 50 mas.
Substantial deviations from the typical pattern from an inclined sub-Keplerian disk are already noticeable in the low inclination case. The following columns show the extracted velocity perturbations, which are seen as quasi axisymmetric rings that reach magnitudes of $\sim 50$ m s$^{-1}$. The sub-structures become more complex as we increase the disk inclination.
In the second column, we present the velocity perturbations obtained when we subtract to v$_0$ shown in the first column, a second velocity centroid map computed from a synthetic observation of a smooth disk following an equilibrium solution (v$_{eq}$). We use here the initial conditions from the simulation as input for the radiative transfer calculations, to compute the synthetic observations of the disk in equilibrium. The resulting residual map can be used as the expected velocity perturbations in an ideal observation with 5 au resolution.
In the third column, the residual of subtracting to v$_0$, in this case computed for a synthetic cube convolved by a 10 au circular Gaussian beam, the best fit disk model obtained with \textsc{eddy} (v$_{mod}$). As detailed in Section \ref{observables}, the model corresponds to a geometrically thick disk, that follows power-law profiles for the velocities in the radial and azimuthal directions. The resulting residual maps are the observable velocity perturbations in a 10 au resolution observation, with the addition of white noise with an RMS level expected for a 20h long-baseline ALMA observation. 
Recovering properly the emission surface and velocity profiles is harder for the highest inclination case presented ($i=35^{\circ}$). We tested a higher inclination case with $i=45^{\circ}$, and we found we were unable to recover the velocity perturbations reliably. The extracted velocity perturbations match fairly well with the expected pattern for inclinations of 5 and 20 degrees. Additional modulations are present depending on the polar angle with respect to the disk's major axis for the inclination of 35 degrees case. These modulations are due to systematic differences between the fitted and the true emission surface, with a secondary contribution from the errors on the disk center, PA, and velocity profiles (see also Figure 12 in \citealt{Yen2020}).
In the fourth column, we applied the same procedure to extract the velocity perturbations shown in the third column, but in this case for a v$_0$ computed from the simulated ALMA observations described in Section \ref{simulatedobs}. We observe that the perturbations are consistent with the expected VSI ringed structure, therefore, VSI-signatures are observable within ALMA capabilities.

How the different velocity components contribute to the projected line of sight velocity (LOSV) is crucial to interpret the velocity deviations. As shown by \citeauthor{Teague2019} 2019 (see their Figure 5), a ring of super(sub) Keplerian azimuthal motions show a sign-flip (from red- to blue-shifted, or vice versa) with respect to the PA of the disk. While for a ring of inward(outward) flow the sign flip is with respect to the line perpendicular to the disk PA. A ring of upward(downward) vertical motion, however, has the same sign for all PA. 
To better understand the contributions of each velocity component to the projected line of sight velocity, we computed the expected velocity perturbations considering each component separately. The synthetic data cubes were calculated using an input velocity field in which only one of the three velocity components is from a VSI unstable disk simulation and the remaining two are set to follow an equilibrium disk solution (v$_{eq}$). In Figure \ref{components} we observe the contributions to the LOSV from the radial velocity $v_r$, meridional velocity $v_{\theta}$ and azimuthal velocity $v_{\phi}$ for different disk inclinations. It is clear that the meridional velocity dominates the LOSV. With increasing inclination, the radial and azimuthal velocities contribute more to the LOSV, where the radial velocity introduces asymmetric features in our highest inclination case. The contribution from the azimuthal velocity perturbations is minor for all the inclinations explored.

To summarize, the meridional flows from the VSI body modes are the most distinctive feature observable in the velocity centroid maps of a VSI-unstable disk. For low disk inclinations we have rings of red- or blue-shifted emission. 
Increasing the inclination we have a significant contribution of the radial velocity, in which the asymmetric nature of the radial perturbations adds complications to interpret the observable VSI signatures. Moreover, modulations of the extracted perturbations depending on the polar angle relative to the disk PA can appear due to the limitations of the method, making the characterization of the VSI-signatures difficult.

\subsection{Detectability of VSI turbulence}

\begin{figure}[htp!]
\centering
\includegraphics[angle=0,width=0.98\linewidth]{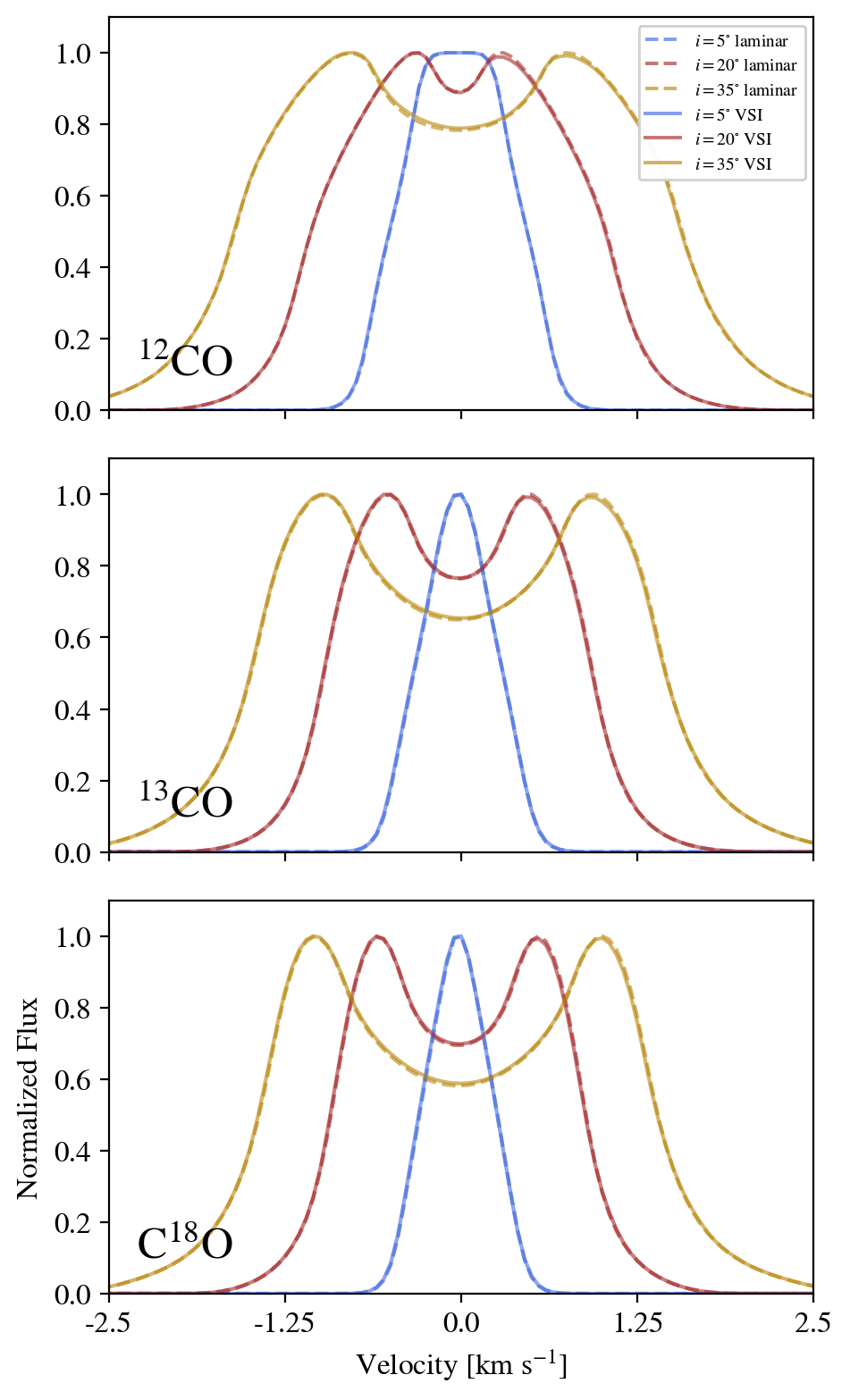}
\caption{Line profiles from synthetic observations for different CO isotopologues and disk inclinations. In each panel we show a comparison of the integrated line emission for a VSI unstable disk (solid lines) and a laminar disk (dashed lines).}
\label{lineprofiles}
\end{figure}

A previous study has shown that a disk with isotropic turbulence can produce observable signatures of non-thermal gas motions in ALMA observations \citep{Simon2015}. 
First, it can produce a change in the shape of the spatially integrated line of emission, increasing the ratio of the peak line flux to the flux at line center. Such diagnosis has been applied to constrain disk turbulence in the outer regions of protoplanetary disks (see e.g. \cite{Flaherty2020} and references therein). We note that \citet{Simon2015} used a parametric fit to non-thermal gas motions to determine a turbulent broadening parameter as a function of radius and height. This parameter was then used as input in the line radiation transfer code LIME for the sake of computational efficiency. For our work we include the full 3D velocity field to calculate the line emission.\\
We explored if we can see the effect of line broadening by turbulence in our simulation data with anisotropic turbulence from the VSI. We compare directly the shape of the line of the synthetic observations for the VSI unstable disk, and a disk that follows a sub-Keplerian equilibrium solution with $v_{turb}=0$, i.e. a laminar disk with the radial and meridional velocities being zero. We assume the same dust temperature structure for both disks. The comparison is shown in Fig. \ref{lineprofiles} for data cubes with spatial resolution of 50 mas (5 au) and velocity resolution of 50 m s$^{-1}$. We observe that the effect of VSI turbulence in the line shape is negligible compared to the thermal broadening for all inclinations and CO isotopologues explored. We recovered similar results for a disk around a central star with lower temperature, analogous to a young Sun, with an effective temperature of $4300$ K and radius of $2.6\,R_{\odot}$. This highlights how important is to use the full 3D velocity data to produce synthetic observation and study the line broadening in protoplanetary disks.
\\ 
The second main result by \citet{Simon2015} showed that isotropic turbulence can affect the distribution of the emission in a given velocity channel. \cite{Flaherty2020} has shown that anisotropic turbulent motions can mimic such effect in spatially or spectrally unresolved observations. We observe that this effect is minor in our predictions, as the physical size and magnitude of the velocity perturbations is rather small (about 10 au and 50 m s$^{-1}$). The finger-like features are already indistinguishable in independent channels for a spatial resolution of about 30 au. 
We summarize again that a full 3D velocity field is needed when studying line-broadening. For $^{12}$CO, the thickness of the emission layer is smaller than a scale height. Furthermore, line broadening is only caused by velocity fluctuations with physical scales smaller than the depth of the emission layer. This fact causes the turbulent broadening to be negligible because the amplitudes of these velocity fluctuations are about one order of magnitude smaller than the large scale fluctuations. We stress that it might be more promising to determine the kinematics by spatially and spectrally resolved observations in contrast to determine the line broadening in unresolved disk observations. 
Our results of the $\alpha$ viscosity values from the hydrodynamical simulations are consistent with the upper limits from non-thermal broadening in ALMA observations, with $\alpha$ viscosity values $\lesssim 10^{-3}$. We compute in our simulation a disk-averaged radial $\alpha$ value of $\alpha_{r\phi}=1.4\times 10^{-4}$, and a value about 3.6 times larger in the vertical direction \citep[computed for one hemisphere of the disk following Eq. 4 in][]{Stoll2017}, i.e. $\alpha_{z\phi} \sim 3.6 \alpha_{r\phi}$. However, we note again that these are not directly comparable as stated above, and going for higher resolution, or more sensitive observations would not result on the constraint of a turbulent $\alpha$ from VSI.
We emphasize that spatially resolving the velocity perturbations from the VSI is required to confirm it as a source of turbulence. Moreover, it is feasible to estimate an $\alpha$ viscosity value of the disk by directly comparing the observed velocity structure with synthetic predictions from 3D hydrodynamical simulations.
We conclude that the currently applied diagnostics to detect non-thermal turbulent motions in ALMA observations using a parametrization of the turbulence level are not reliable if the turbulence is anisotropic. Therefore, we can not discard that the VSI is active in these disks, and that turbulence is still playing an important role.
Finally, despite our simulations do not resolve the smallest scales of the instability \citep{Flores2020}, we do not expect a change in the turbulent line broadening when resolving these. The smaller turbulence scales are expected to have lower RMS velocities due to the turbulent cascade. Therefore, they do not significantly contribute to the broadening of the integrated line emission.

\section{Discussion}\label{discussion}

\subsection{Effect of optical depth}

Observing different CO isotopologues, allows to study different layers of the disk.
To test the effect of the tracer optical depth, we run the radiative transfer calculations for $^{12}$CO, $^{13}$CO and C$^{18}$O. The disk layers traced approximately by the different isotopologues are shown in Figure \ref{layers}, where we display the layer where the disk becomes optically thick ($\tau=1$ surface). For simplicity, these calculations are for a disk face-on.
The emissions traced in our predictions roughly match with constraints from observations \citep{Zhang2017, Pinte2018b}.
The expected velocity perturbations from a VSI-unstable disk for the three different isotopologues is shown in Figure \ref{tracers}, for different disk inclinations.\\
We observe that the quasi axisymmetric ring structure present in the residuals of the $^{12}$CO predictions are also recovered for the $^{13}$CO and C$^{18}$O predictions. Further, the remaining velocity residuals are feebler for optically thinner tracers, where the difference between isotopologues becomes more noticeable with increasing disk inclination. 
These findings are expected since the emission layer for these lines trace lower disk heights where the VSI perturbations are weaker. Moreover, the radial and azimuthal velocity components decrease faster than the vertical moving towards the midplane, as seen in the $r$-$\theta$ slices of the disk in Figure \ref{fighydrosimus_all}, where is shown that the vertical motions from the VSI body modes penetrate through the whole vertical extent of the disk. 
Therefore, the meridional component of the velocity perturbations is predicted to also dominate the line of sight velocity at deeper layers, resulting in the recovered quasi axisymmetric ringed morphology (as discussed in Section \ref{Deviations}). Furthermore, the difference among tracers becomes larger for more inclined disks as the contribution to the line of sight velocity from the radial and azimuthal gas velocity increases with inclination.
Additionally, the velocity structure is smeared for optically thinner tracers due to the emission arising from a larger range of heights \citep{Zhang2017}.
Therefore, detecting VSI signatures in $^{13}$CO or C$^{18}$O would require higher sensitivity, especially for inclined disks. Adding the difficulties to get high spatial resolution observations with enough signal-to-noise, we conclude that $^{12}$CO is the best tracer to detect VSI signatures. Nevertheless, observing the weak dependence with disk height of the morphology and magnitude of the VSI velocity perturbations for low inclined disks is essential to confirm its origin.

\subsection{How to differenciate VSI signatures from other mechanisms?}\label{differenciate}

Recognizing VSI-driven gas motions is very important to interpret kinematics observations. Large-scale perturbations in the gas velocities can also be produced by other mechanisms. For example, massive planets \citep{DiskDynamics2020}, vortices \citep{Huang2018, Robert2020} or gravitational instabilities \citep{Hall2020}. 
A few points that can be used to separate the VSI signatures from other mechanisms are the following:
\begin{itemize}
    \item VSI signatures are quasi axisymmetric. Velocity perturbations of similar magnitude are visible in all velocity channels (see Figure \ref{channelmaps}), and show ringed sub-structures when extracted from the Moment one map (see Figure \ref{observables}). On the contrary, perturbations from the aforementioned alternative mechanisms are asymmetric, as trace spiral-arms or localized deviations. Planetary-wakes and vortices are expected to be strongest in a few velocity channels and show a localized signature in the velocity centroid maps \citep{Perez2015, Perez2018, Huang2018, Robert2020}. Large-scale spiral arms, however, can also appear across all velocity channels, yet are distinguishable from rings when extracted from the velocity centroid maps \citep{Perez2018,Hall2020}.
    \item The velocity component that contributes more to the line of sight velocity of the perturbations in a VSI-unstable disk is the vertical component. For planetary wakes and spiral-arm structures the largest contributions are from the radial and azimuthal velocities \citep{Perez2018, Hall2020}. Meridional flows can also be generated in the gaps carved by embedded massive planets \citep{Kley2001, Morbidelli2014, Fung2016, Teague2019}. For Jupiter mass planets, the azimuthally averaged velocities at the gap and gap's edges can have similar magnitude as the predicted VSI signatures \citep{Teague2019}. However, such flows are typically present together with spiral arms and perturbations localized around the planet's location \citep{Perez2018, Juhasz2018}. Therefore, resolved two-dimensional velocity centroid maps are required to disentangle between both scenarios.
    \item The vertical motions of the VSI-modes are present in the whole disk vertical extent. Therefore, no significant change in the morphology of the observable VSI signatures are predicted when tracing different CO isotopologues. For planetary wakes and spiral arms, it is expected that the observable velocity perturbations change depending on disk height \citep{Perez2015}.
    \item An embedded planet in a slow-cooling disk can generate additional tightly wound spiral arms due to buoyancy resonances. These spirals can have a strong vertical velocity component and produce observational signatures that can be confused with VSI signatures \citep{Bae2021}. Nevertheless, high spatial resolution observations can potentially disentangle between tightly wound spirals and VSI quasi axisymmetric rings in the line of sight velocity residuals.
    \item VSI-perturbations are expected to produce dust traffic jams in the outer disk \citep{Flock2017, Flock2020,schaefer2020}. It can also generate a dust trap at the radial transition where the cooling-timescale is short enough to sustain the VSI \citep{Flock2020}. Additionally, it can trigger Rossby Wave Instability vortices \citep{Flock2020,Manger2018}, which can efficiently trap dust at its center \citep{BargeSommeria1995}. Similar structures in the dust distribution can be produced by embedded planets. Thus, it is difficult to disentangle between VSI and other mechanisms when studying the radial and azimuthal distribution of the dust continuum. Nevertheless, by studying the vertical distribution of the dust from edge on it is possible to assess if dust vertical mixing from the VSI is present \citep{Flock2017, Flock2020, Villenave2020}.
\end{itemize}

In our simulation, short-lived small-scale Rossby Wave Instability vortices appear, consistent with previous results on VSI-unstable disks with very short thermal relaxation time scales \citep{Richard2016, Flock2020}. These vortices could produce kinematic signatures in CO line emission, as previously shown for large-scale planet-induced vortices \citep{Huang2018, Robert2020}. However, we focus on the global quasi-axisymmetric ringed structure produced by the VSI. Further analysis is needed to constrain the observability of vortices induced by the VSI in CO kinematics.

\subsection{Constraints on the disk properties from a VSI-detection}

Strong constraints can be done from the detection of the VSI in a protoplanetary disk. First, a VSI detection would confirm that the disk has a vertical shear and a radial gradient of the temperature.
Second, a short cooling timescale is necessary for the VSI to operate in the traced layers of the disk. Via hydrodynamical simulations it has been shown that the VSI is active in disks with cooling timescales $<10\%$ of the local orbital timescale \citep{Nelson2013, Lin2015, Flock2020}. Hence, a VSI detection would constrain the disk cooling timescales.
Finally, it also shows that the turbulence in the disk is fully or partly due to the VSI, which has been constrained from hydrodynamical simulations to $\alpha$ viscosity values in the range of $10^{-5}$ to $10^{-3}$.
Damping of the VSI is predicted for the upper layers of the disk where the densities are low enough to allow collisional dust-gas decoupling \citep{Pfeil2020} and magnetic effects dominate \citep{Cui2020}. Therefore, studying the velocity perturbations at different layers of the disk can help us to understand changes in physical conditions at different disk heights.

\subsection{Observations spatial and spectral resolution}

We explored the effect of spatial and spectral resolution on the observable velocity perturbations for a disk with an inclination of 20 degrees. We display in Figure \ref{resolutions} the expected velocity perturbations from a VSI unstable disk varying the velocity resolution of the data cube, and size of the convolved circular Gaussian beam, in the first and third row, respectively. Additionally, the statistical uncertainties from the calculation of the line centroid maps derived using \textsc{eddy} \citep{Teague2018c} are shown in the third and fourth rows. Each data cube has added white noise with an RMS level expected from a 20 h integration ALMA observation.
We observe that for beam sizes $\leq 10$ au the VSI structure is well resolved. For such a high spatial resolution, all velocity resolutions explored reach an uncertainty level smaller than $\sim 20\%$ the channel width and the structure is recovered. However, only for the highest velocity resolution cases (channel width $\leq 50$ m s$^{-1}$) these uncertainties are below $\sim 20\%$ the magnitude of the velocity perturbations. 
In summary, we found that high resolution observations using ALMA extended antenna configurations can spatially resolve the VSI signatures. However, the highest spectral resolution observations are needed for a robust detection. 

For our predictions, with an assumed distance to the system of 100 pc, a 0.1 arcsecond resolution is enough. In ALMA band 6, the antenna configuration 8 gives the required resolution, where a 20h integration time can get sufficient signal-to-noise ratio. Disks further away would require higher angular resolution to resolve the VSI, which would require a longer integration time. 
Our simulated observation setup is at the limit of the capabilities available for ALMA Cycle 8. Therefore, detection of VSI-signatures in the next ALMA cycle might be feasible for the brightest protoplanetary disks only.\\
Considering a standard Hanning smoothing, the highest resolutions are $\sim 0.05$ km s$^{-1}$ and$\sim0.03$ km s$^{-1}$, for ALMA Bands 6 and 7, respectively. According to our predictions, with these resolutions it is possible to resolve well the VSI signatures. Nonetheless, no spectral averaging and avoiding Hanning smoothing when preparing ALMA observations allows to achieve finer spectral resolution (down to $\sim 0.025$ km s$^{-1}$ for Band 7), which could ease the identification of VSI signatures.\\
We highlight that a good calibration is key to recover velocity perturbations from the data. For such a high spectral resolution observation to be successful, at least one of the basebands should be set up in TDM/wideband to get the best calibration. Otherwise, there may not be enough total bandwidth for best phase calibration, and this is truly essential to get the highest dynamic range images using self-calibration.\\ 
Our predictions are also valid for the CO J:3-2 transitions, observable within ALMA Band 7 in which higher velocity resolution observations are possible. Moreover, observing $^{12}$CO and $^{13}$CO within the same observation can be done for the J:3-2 transition. Obtaining the VSI signatures at different layers of the disk is important to better understand the vertical flows in the disk, and can potentially confirm the origin of the velocity perturbations.\\
The velocity resolution required to identify the VSI signatures could also vary depending on the method used to compute the line centroids and extract the velocity perturbations. Alternative methods to obtain the moment maps and the non-perturbed velocity centroid map are \textsc{GMoments\footnote{\url{https://github.com/simoncasassus/GMoments}}}, and \textsc{ConeRot\footnote{\url{https://github.com/simoncasassus/ConeRot}}} (\citealt{Casassus2019}, Casassus et al. in prep). The advantage of \textsc{ConeRot} compared to \textsc{Eddy} is that the perturbations can be extracted directly from observations without strong assumptions about the underlying disk model, and employing a reduced number of free parameters. Nevertheless, we do not expect significant differences in our results, but in its application to ALMA data.\\
Last, further study of the disk velocity structure can be performed analysing the velocity radial profiles. For this purpose \textsc{ConeRot} and \textsc{gofish\footnote{\url{https://github.com/richteague/gofish}}}\citep{GoFish} can be applied. However, as discussed in Section \ref{differenciate}, a thorough study of the three velocity components at different layers of the disk would be required to possibly disentangle between perturbations produced by planets or VSI.

\begin{table}
\caption[Simulation parameters]{Summary of simulation and radiative transfer parameters.} 

	\centering

	\begin{tabular}{cc}
	\hline
    \hline
    	Parameter & Value \\
        \hline
        Reference radius & 100 au \\
        Aspect ratio at 100 au  & 0.1 \\
        Flaring index  & 0.25 \\
        Surface density slope & -1.0 \\
        Temperature slope & -0.5  \\
        Stellar mass & 1.0 $M_{\odot}$  \\
        $\#$ of cells in $r$ & 512   \\
        $\#$ of cells in $\phi$ & 1024 \\
        $\#$ of cells in ${\theta}$ & 128   \\
        Grid inner radius & 40 au  \\
        Grid outer radius & 250 au  \\
        Total evolution time & 0.3 Myr  \\
        \hline
        Stellar radius & $ 1.0\,R_{\odot}$\\
        Stellar effective temperature & $7000\,\textrm{K}$\\
        Distance & 100 pc \\
        Disk total mass & 0.05 $M_{\star}$  \\
        Dust-to-gas mass ratio  & 10$^{-2}$  \\
        Maximum dust size & 1 cm  \\
        Minimum Dust Size & 0.01 $\mu$m  \\
        Dust size slope & -3.5  \\
        Dust intrinsic density & 2.7 g cm$^{-3}$  \\
        Disk inclination & [$5^{\circ}$,$20^{\circ}$,$35^{\circ}$] \\
        \hline
        \hline
	\end{tabular} 
	
\label{tab:simparams}
\end{table}

\section{Summary}\label{conclusions}

We conducted a study of the observability of gas motions produced by the vertical shear instability (VSI) in ALMA gas kinematics observations. We explore the effect of disk inclination, spatial and spectral resolution. Furthermore, we compare the results for different CO isotopologues.
We couple the results of 3D global hydrodynamical simulations of a disk unstable to the VSI, with radiative transfer calculations to obtain synthetic line emission predictions for $^{12}$CO, $^{13}$CO and C$^{18}$O molecules. Next, we produce simulated observations that are directly comparable to those obtained by ALMA. Finally, we extract the velocity perturbations from our simulated observations, and recover the VSI perturbations. Our main findings are:

\begin{itemize}

    \item We find that the VSI perturbations are seen as a corrugated pattern in the emission of individual velocity channels, and quasi axisymmetric concentric rings in the line of sight velocity residuals after subtracting a sub-Keplerian disk model fitted to the data. 
    
    \item The characteristic morphology of the extracted perturbations results from the meridional velocity component being dominant in the velocity projected into the line of sight.

    \item With increasing inclination, the line of sight velocity has stronger addition of the radial and azimuthal velocity of the disk, adding asymmetries on top of the rings, making the VSI-perturbations challenging to characterize. Moreover, the method to extract the perturbations does not correctly recover the expected morphology for large inclinations ($i\gtrsim 35^{\circ}$).
 
    \item The morphology of the extracted VSI velocity perturbations is predicted to be similar for different layers of the disk. The change in the magnitude of the expected velocity perturbations for different tracers remains relatively low for all inclinations explored. Further, the difference in magnitude increases with disk inclination. 
    
    \item Our results show that the non-thermal broadening produced by the VSI in integrated line emission is negligible, consistent with current limits from ALMA radio observations. We emphasize that resolving the structure is fundamental to determine the turbulence sources in the disk.

    \item To recover the VSI-perturbations, spatial resolution below the disk pressure scale height might be enough to resolve the structures, which in our case corresponds to beam sizes $\lesssim 10$ au. However, the spectral resolution needed to capture the perturbations reliably ($\lesssim 0.05\,\rm\,km\,s^{-1}$) is the highest available with the ALMA interferometer. 
 
\end{itemize}

We highlight that our predictions are optimistic and the predicted observational signatures should be treated as upper limits. The conducted simulations are inviscid and follow a locally isothermal equation of state, giving the ideal conditions for the development of vigorous VSI motions in the disk. The inclusion of a finite cooling time or background viscosity can weaken the VSI perturbations. Moreover, the presence of velocity perturbations from a different origin (e.g. massive planets) can obstruct the identification of the VSI signatures. Further work is needed to study these scenarios. \\
Observing the gas kinematics at different layers of the disk is important to disentangle between the VSI and other mechanisms that could produce similar kinematic signatures. Finally, we conclude that the best cases to detect VSI signatures are gap-less disks close to face-on, in which the VSI meridional flows can be directly traced.

\newpage
\appendix

\section{Appendix material}

\begin{figure*}[htp!]
\centering
\includegraphics[angle=0,width=0.9\linewidth]{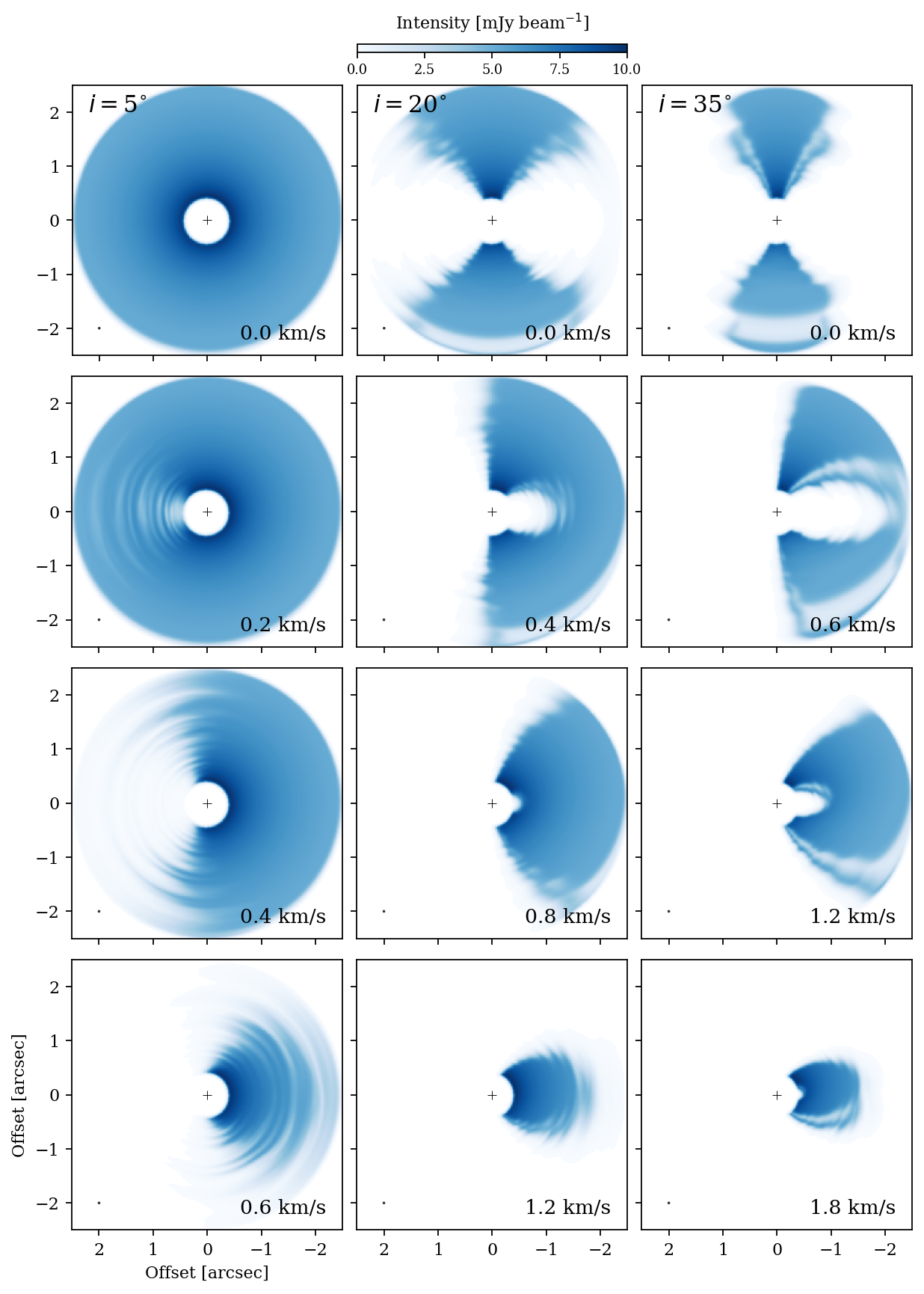}
\caption{Predictions of the $^{12}$CO disk emission of a VSI-unstable disk for different velocity channels.
From left to right, the columns correspond to predictions for disk inclinations $i=5$, $20$ and $35$ degrees. The channel widths are $0.05$ km s$^{-1}$. The images are convolved with a circular Gaussian beam of 50 mas, shown in the bottom-left corner of each panel.
The disk emission shows a corrugated pattern. Only red-shifted channels are shown, as the structures are quasi axisymmetric, therefore, the symmetric blue-shifted channels show similar sub-structures. }
\label{channelmaps}
\end{figure*}

\begin{figure*}[htp!]
\centering
\includegraphics[angle=0,width=1.0\linewidth]{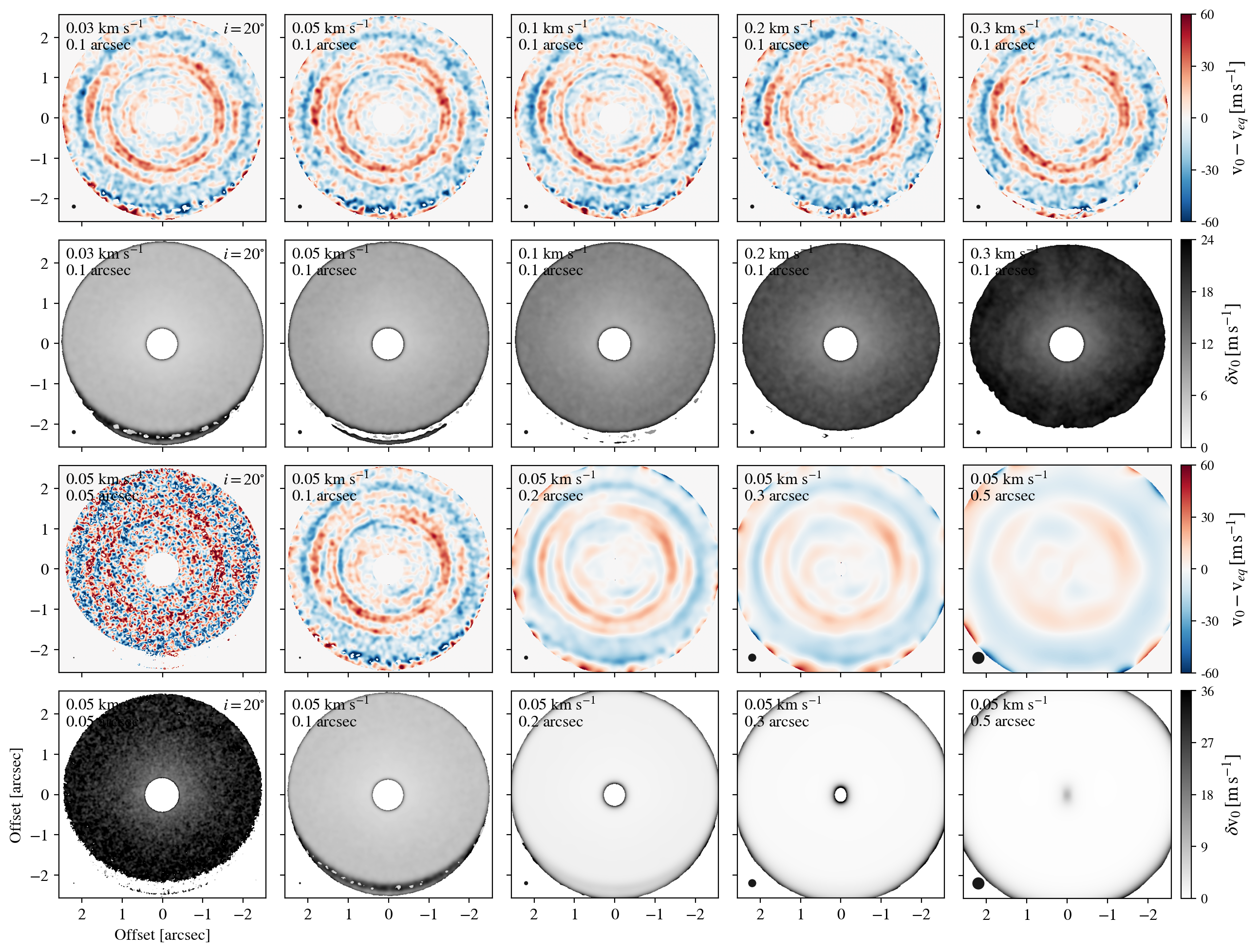}

\caption{Results of the expected observable velocity perturbations from a VSI unstable disk $^{12}$CO synthetic lines observations ($v_0-v_{eq}$), for various velocity and spatial resolution data cubes. Each channel of the cube has a RMS noise expected from a 20 h integration ALMA observation with an extended configuration (1.9, 1.5, 1.0, 0.8, 0.6 mJy from the higher to the lower velocity resolution cases).
In addition, we show the uncertainties in the calculation of the velocity centroid map using \textsc{bettermoments} ($\delta v_0$).
The FWHM of the circular Gaussian beams are shown at the bottom left corner of each panel. The x- and y- axes indicate the R.A. and Dec. angular offset from the position of the central star, in arcseconds.}
\label{resolutions}
\end{figure*}

\begin{figure*}[htp!]
\centering
\includegraphics[angle=0,width=1.0\linewidth]{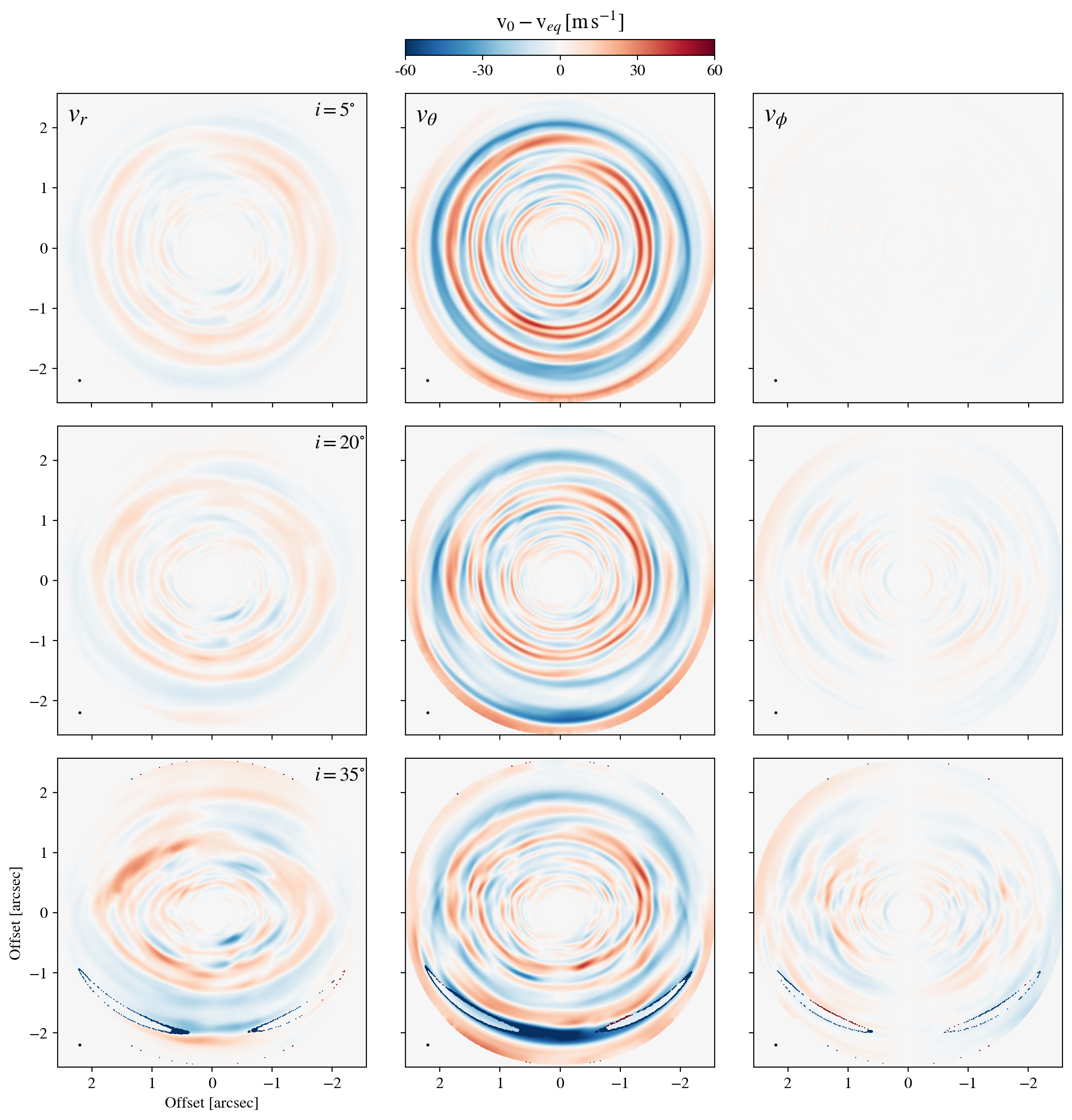}
\caption{Results of the expected observable velocity perturbations from $^{12}$CO synthetic lines observations, in which one of the velocity components is from a VSI unstable disk. The input VSI-active velocity field into the radiative transfer is modified to follow an equilibrium disk solution in two of the components.
From left to right, the columns show predictions considering the contributions of the radial velocity, the meridional velocity and azimuthal velocity, respectively. From top to bottom, the different rows show the results for disk inclinations of $5^{\circ}$, $20^{\circ}$ and $35^{\circ}$.
The mock data cubes have a velocity resolution of 0.05 km s$^{-1}$, and the images are convolved by a circular Gaussian beam of 50 mas, shown at the bottom left corner of each panel. 
The x- and y- axes indicate the R.A. and Dec. angular offset from the position of the central star, in arcseconds.}
\label{components}
\end{figure*}

\begin{figure*}[htp!]
\centering
\includegraphics[angle=0,width=1.0\linewidth]{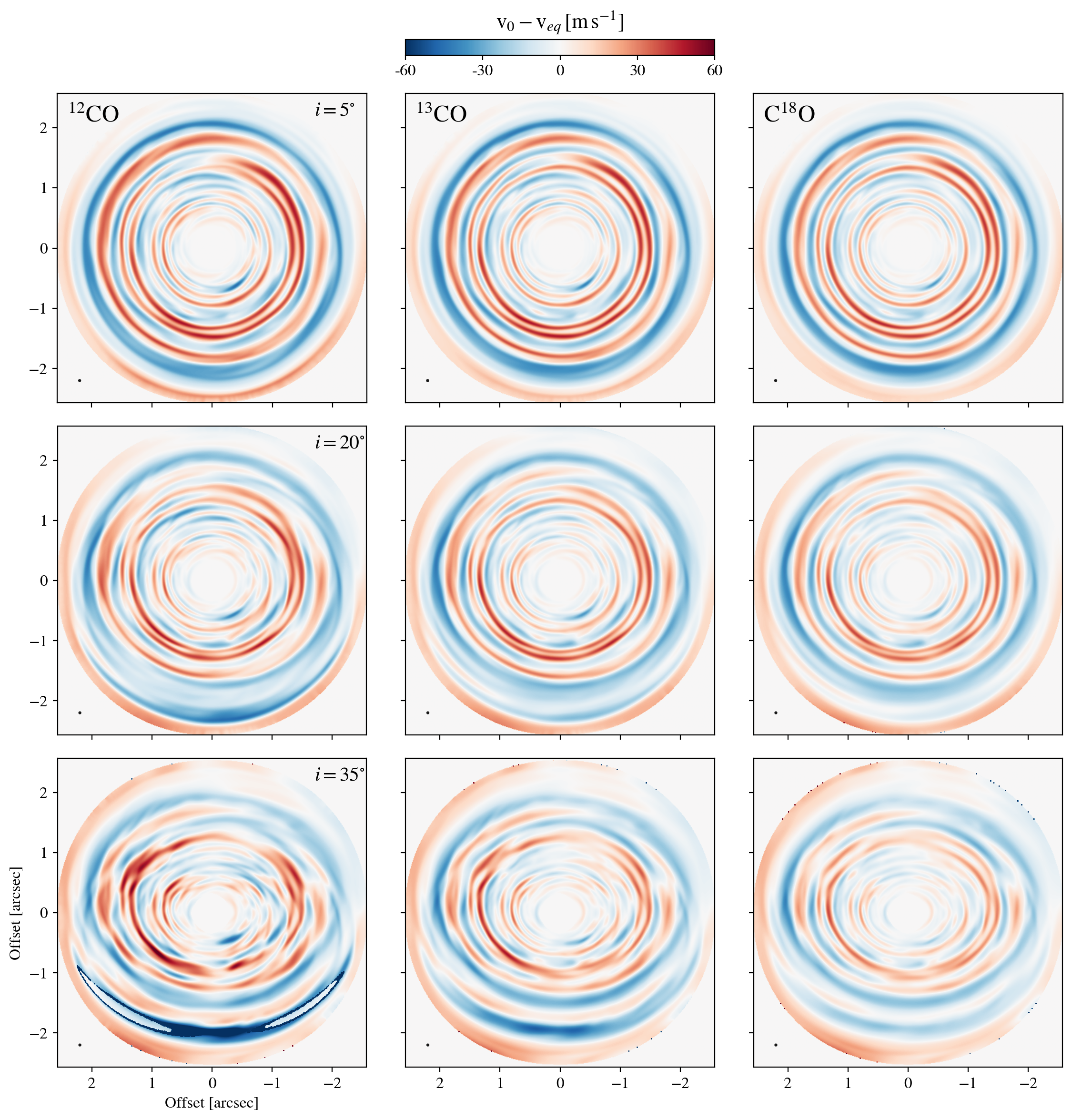}
\caption{Results of the expected observable velocity perturbations from a VSI unstable disk synthetic lines observations for different CO isotopologues. From left to right, the columns show predictions for $^{12}$CO, $^{13}$CO and C$^{18}$O. From top to bottom, the rows show the results for disk inclinations of $5^{\circ}$, $20^{\circ}$ and $35^{\circ}$. The mock data cubes have a velocity resolution of 0.05 km s$^{-1}$, and the images are convolved by a circular Gaussian beam of 50 mas, shown at the bottom left corner of each panel.
The x- and y- axes indicate the R.A. and Dec. angular offset from the position of the central star, in arcseconds.}
\label{tracers}
\end{figure*}

\begin{figure*}[htp!]
\centering
\includegraphics[angle=0,width=1.0\linewidth]{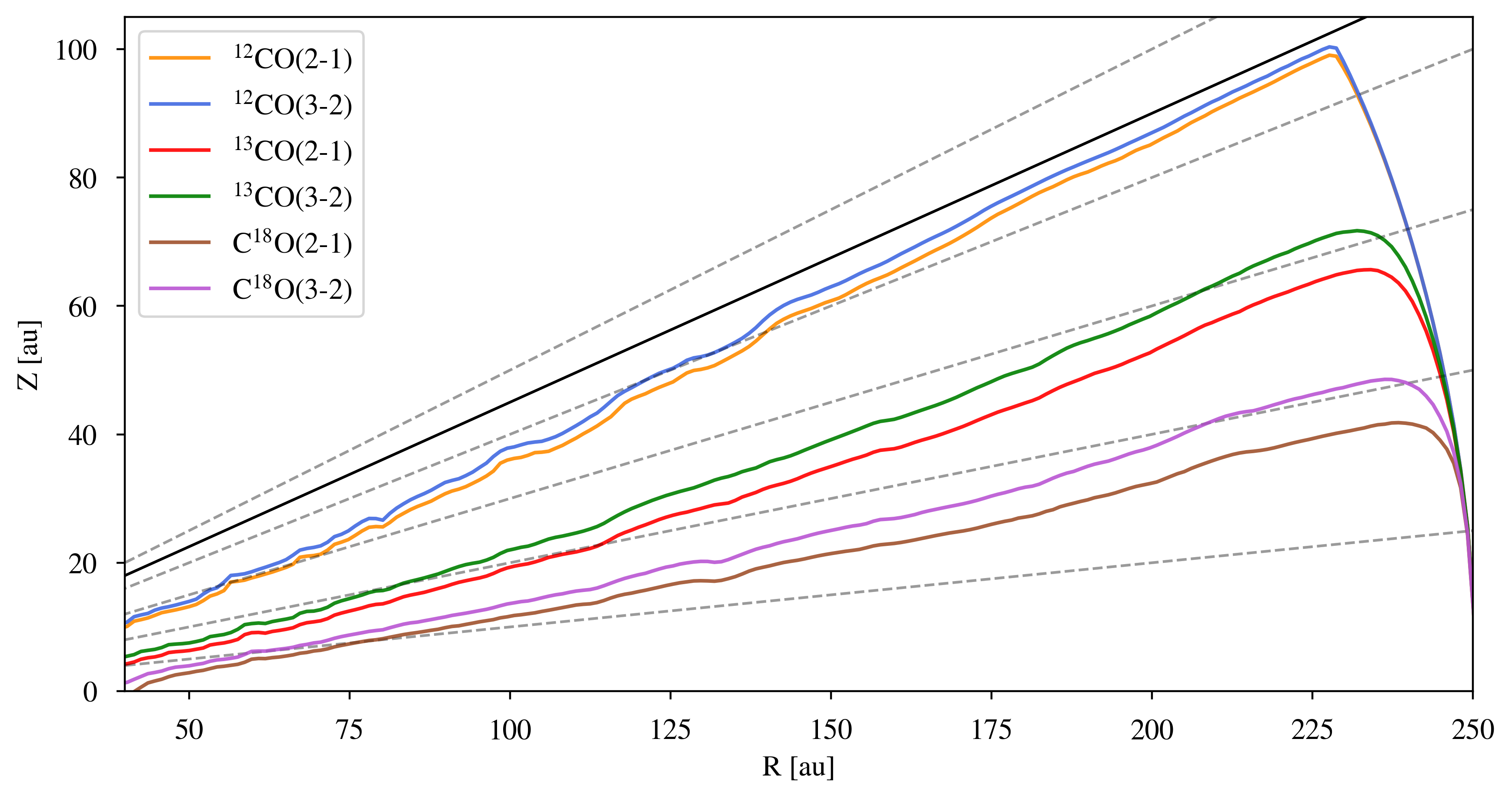}
\caption{Radial profile of the disk layers where the disk becomes optically thick ($\tau=1$ surface), for different CO isotopologues of the VSI-unstable disk radiative transfer model. We illustrate the $\tau=1$ surface for the J:2-1 and J:3-2 transitions of three CO isotopologues, $^{12}$CO, $^{13}$CO and C$^{18}$O. We assume a face-on disk orientation for the calculation. The grey dashed lines show where $Z=$A$R$ for A$=0.1;0.2;0.3;0.4$ and $0.5$. The solid black line shows the grid boundary in colatitude of the simulated disk, located approximately at $Z=0.45R$.}
\label{layers}
\end{figure*}

\begin{acknowledgements}
We thank the anonymous referee for constructive and detailed review of the manuscript. We thank F. Alarc\'on, N. Kurtovic and R. Teague, for helpful advice on the use of RADMC3D, CASA and bettermoments/eddy, respectively.
Figures where produced by python matplotlib library \citep{Hunter:2007}, and ParaView \citep{Paraview1, Paraview2}. M.B and M.F. acknowledges funding from the European Research Council (ERC) under the European Union’s Horizon 2020 research and innovation program (grant agreement No. 757957). S.M. is supported by a Junior Research Fellowship from Jesus College, Cambridge. S.P. acknowledges support from ANID-FONDECYT grant 1191934. The numerical simulations were run on the HPC system COBRA at MPCDF (Max Planck Computing and Data Facility).
      
\end{acknowledgements}

%
%


\bibliographystyle{aa} 
\bibliography{bibliography} 

\end{document}